\documentstyle[epsfig,rotate,12pt]{article}
\begin{document}
\renewcommand{\thefootnote}{\arabic{footnote}}
\setcounter{footnote}{0}
\newcommand{\go}[1]{\gamma^{#1}}
\newcommand{\gu}[1]{\gamma_{#1}}
\newcommand{\etsl}{\eta \hspace{-6pt} / }
\newcommand{\epsl}{\epsilon \hspace{-5pt} / }
\newcommand{\qsl}{q \hspace{-5pt} / }
\newcommand{\pbsl}{p \hspace{-5pt} / }
\newcommand{\qssl}{q' \hspace{-7pt} / }
\newcommand{\pssl}{p' \hspace{-7pt} / }
\newcommand{\ra}{\rightarrow}
\newcommand{\Li}{\mbox{Li}}
\newcommand{\BR}{{\cal B}}
\newcommand{\M}{{\cal M}}
\newcommand{\BGAMAXS}{B \ra X _{s} + \gamma}
\newcommand{\BGAMAXD}{B \ra X _{d} + \gamma}
\newcommand{\BBGAMAXS}{\BR (B \ra  X _{s} + \gamma)}
\newcommand{\BBGAMAXD}{\BR (B \ra  X _{d} + \gamma)}
\newcommand{\BBKSTAR}{\BR (B \ra K^\star + \gamma)}
\newcommand{\BKSTAR}{B \ra K^\star + \gamma}
\newcommand{\GGAMAXS}{\Gamma (B \ra  X _{s} + \gamma)}
\newcommand{\GGAMAXD}{\Gamma(B \ra  X _{d} + \gamma)}
\newcommand{\bsgam}{\ $b \to s+ \gamma$}
\newcommand{\bdgam}{\ $b \to d+ \gamma$}
\newcommand{\bsg}{\ $b \to s+ g$}
\newcommand{\bsggam}{\ $b \to s+ \gamma+ g$}
\newcommand{\bdggam}{\ $b \to d+ \gamma+ g$}
\newcommand{\aqcdr}{\frac{\alpha_s}{3 \pi}}
\newcommand{\bgamaxs}{$B \to X _{s} + \gamma$}
\newcommand{\bgamaxd}{$B \to X _{d} + \gamma$}
\newcommand{\brbgamaxd}{{\cal B}(B \to X _{d} + \gamma)}
\newcommand{\bksgam}{\ $B \to K^*+ \gamma$}
\def\BDSl{B \to D^* + \ell \nu_\ell}
\newcommand{\Tosc}{T_{osc}}
\def\Vbc{V_{cb}}
\def\Vbu{V_{ub}}
\def\Vtd{V_{td}}
\def\Vts{V_{ts}}
\def\Vtb{V_{tb}}
\newcommand{\absvcb}{\vert V_{cb}\vert}
\newcommand{\absvub}{\vert V_{ub}\vert}
\newcommand{\absvtd}{\vert V_{td}\vert}
\newcommand{\absvts}{\vert V_{ts}\vert}
\newcommand{\xd}{x_d}
\newcommand{\bd}{B_d^0}
\newcommand{\mt}{m_t}
\newcommand{\mb}{m_b}
\newcommand{\mc}{m_c}
\newcommand{\bbar}{B^0-\bar {B^0}}
\newcommand{\bdb}{\overline{B_d^0}}
\newcommand{\delmd}{\Delta M_d}
\newcommand{\delms}{\Delta M_s}
\newcommand{\ps}{10^{-12} s}
\newcommand{\kkbar}{$K^0$-${\overline{K^0}}$}
\newcommand{\bdbdbar}{$B_d^0$-${\overline{B_d^0}}$}
\newcommand{\bsbsbar}{$B_s^0$-${\overline{B_s^0}}$}
\newcommand{\abseps}{\vert\epsilon\vert}
\newcommand{\fbb}{f^2_{B_d}B_{B_d}}
\newcommand{\epsp}{\epsilon^\prime/\epsilon}
\newcommand{\fbbs}{f^2_{B_s}B_{B_s}}
\newcommand{\fbd}{f_{B_d}}
\newcommand{\fbs}{f_{B_s}}
\newcommand{\fds}{f_{D_s}}
\def\rts{\sqrt{s}}
\newcommand{\ba}{\begin{array}}
\newcommand{\ea}{\end{array}}
\newcommand{\be}{\begin{equation}}
\newcommand{\ee}{\end{equation}}
\newcommand{\bea}{\begin{eqnarray}}
\newcommand{\eea}{\end{eqnarray}}
%
\def\qb{\bar{q}}
\def\ub{\bar{u}}
\def\db{\bar{d}}
\def\cb{\bar{c}}
\def\sb{\bar{s}}
%
\def\bra{\langle}
\def\ket{\rangle}
%
\def\a{\alpha}
\def\b{\beta}
\def\a{\alpha}
\def\b{\beta}
\def\g{\gamma}
\def\d{\delta}
\def\e{\epsilon}
\def\p{\pi}
\def\ve{\varepsilon}
\def\ep{\varepsilon}
\def\et{\eta}
\def\l{\lambda}
\def\m{\mu}
\def\n{\nu}
\def\G{\Gamma}
\def\D{\Delta}
\def\L{\Lambda}
\def\to{\rightarrow}
\def\pol{\epsilon}
\def\ms{\hat m_s}
\def\u{\hat w(\hat s)}
\def\s{\hat s}
\def\QA{(1-\ms^2)^2-(\s^2+\u^2 z^2)}
\def\QB{-{{1+\ms^2}\over 2}(\s^2-\u^2 z^2 -2 (1+\ms^2) \s+(1-\ms^2)^2)}
\def\QE{-(1+\ms^4+6\ms^2) \s +(1+\ms^2)(1-\ms^2)^2}
\def\QC{(\ms^2+1) \s-(1-\ms^2)^2}
\def\QD{(1+\ms^2) \u z}
\def\PCO{{\cal B}(B \rightarrow l {\bar\nu} +X) \over
   \vert V_{cb} \vert ^2 g(m_c/m_b)}
\def\PA{-2\s^2+\s(1+\ms^2)+(1-\ms^2)^2}
\def\PB{-(1+\ms^2)\s^2-(1+14\ms^2+\ms^4)\s+
2(1+\ms^2)(1-\ms^2)^2}
\def\PC{(1+\ms^2)\s-(1-\ms^2)^2}
\begin{titlepage}
\begin{flushright}
DESY 95-157\\
hep-ph/yyyyy\\
August 1995
\end{flushright}
\vspace*{1.0cm}
\begin{center}
\Large\bf Rare Radiative $B$ Decays in the Standard Model
\end{center}

\vspace{1.0cm}

\begin{center}
Ahmed Ali\\
{\sl Deutsches Elektronen-Synchrotron DESY, D-22603 Hamburg}
\end{center}

\vspace{1.2cm}
\begin{abstract}
\noindent
A status report on the theory and phenomenology of rare radiative $B$
decays
in the standard model is presented with emphasis on the measured
 decays $B \to X_s  \gamma$ and $B \to K^*  \gamma$. Standard model is
 in agreement with experiments though this
comparison is not completely quantitative due to imprecise data
and lack of the complete next-to-leading order contributions to the
decay rates.
Despite this, it is possible to extract non-perturbative parameters
from the shape of the photon energy spectrum in $B \to X_s \gamma$,
such as the $b$-quark mass and the kinetic energy of the $b$ quark in
$B$ hadron. The measured decay rate $\BBGAMAXS =(2.32\pm 0.67) \times
10^{-4}$ can also be used to extract the CKM ratio $\absvts/\absvcb$,
yielding  $\absvts/\absvcb =1.10
\pm 0.43$. Issues bearing on the determination of the parameters of
the CKM matrix from the CKM-suppressed decays $B \to X_d + \gamma$
and $B \to (\rho,\omega) + \gamma$ are also discussed. It is argued that
valuable and independent constraints on the CKM matrix can be obtained
from the measurements of these decays, in particular those
involving neutral $B$-mesons.
\end{abstract}

\vspace{1.0cm}

\centerline{\it to appear in the Proceedings of the XXXth Rencontres
de Moriond}
\centerline{\it ``Electroweak Interactions and Unified Theories''}
\centerline{\it Les Arcs, France, March 1995}

\end{titlepage}

\section{Introduction}
\indent
   The last two years have seen the first observations of the
electromagnetic
penguins in $B$ decays by the CLEO collaboration.
These include the measurements of the exclusive decay rate,
$\BBKSTAR = (4.5 \pm 1.0\pm 0.9) \times 10^{-5}$ \cite{CLEOrare1}, and the
inclusive rate $\BBGAMAXS =(2.32 \pm 0.67) \times 10^{-4}$ \cite{CLEOrare2},
yielding  $R(K^*/X_s) \equiv \Gamma (\BKSTAR )/\GGAMAXS =0.19 \pm 0.09$.
 In addition, the charged and neutral
$B$-meson decay rates are found equal within experimental measurements.
 The inclusive decay rate is
 in agreement with the predictions of
the standard model \cite{ag1,Ciuchini,Buras94}, with the rate
estimated as $\BBGAMAXS = (2.55 \pm 1.28) \times 10^{-4}$
 assuming $\absvts/\absvcb=1.0$ \cite{ag95}. Conversely, one can vary
the Cabibbo-Kobayashi-Maskawa
 (CKM) matrix element ratio and determine it from $\BBGAMAXS$,
which yields $\absvts/\absvcb=1.10 \pm 0.43$.
 The ratio $R(K^*/X_s)$
 is well explained by the QCD-sum-rule
based estimates of the recent vintage \cite{abs93,bksnsr} and by
 wave-function models combined with vector meson dominance
(local parton-hadron duality) \cite{ag1}.
This and the near equality of the charged and neutral decay rates imply
that the observed radiative $B$ decays are dominated by the (common)
electromagnetic penguin (short distance) amplitudes and the
 contributions from $B$-meson-specific diagrams (weak annihilation,
$W^\pm$-exchange) are small.

\indent
The photon energy spectrum in $\BGAMAXS$
 yields information on the structure function of the photon in the
electromagnetic penguins \cite{neubertbsg,Bigietal,KS94,klp95}.
In specific models \cite{Alipiet,shifmangamma}, this information can be
transcribed in terms of non-perturbative
parameters, such as the $b$-quark mass and the kinetic energy of the
$b$ quark in $B$ hadron. These quantities can also be estimated in the
framework of the heavy quark effective theory
\cite{HQETpower} combined with  QCD
sum rules \cite{bqmass,BB94}.
The results of a recent analysis of the CLEO data on $\BGAMAXS$ \cite{ag95}
are consistent with the values expected from such theoretical
considerations.
However, this agreement is presently not completely quantitative due to
imprecise data.

There is considerable interest in measuring the
CKM-suppressed radiative $B$ decays, such as $\BGAMAXD$ \cite{ag2}
and $B \to (\rho,\omega) + \gamma$ \cite{abs93}. A
determination of the CKM parameters from eventual measurements of these
decays requires careful treatment of the competing short-distance (SD) and
long-distance (LD) effects.
This problem  can be formulated in terms of
model-independent correlation functions involving matrix elements of a few
dimension-6 operators in an effective theory.
Techniques, such as the QCD sum rules, can
then be invoked to estimate them. In
\cite{ab95,wyler95}, the leading
LD-effects in the exclusive decays $B \to (\rho, \omega) + \gamma$ are
calculated in terms of the weak annihilation amplitudes.
The largest such effects may show themselves in the charged $B^\pm$-decays,
$B^\pm \to \rho^\pm + \gamma$, contributing up to $O(15 \%)$ of the
corresponding SD-amplitudes; their influence in
the neutral $B$-decays is estimated to be much smaller. Hence,
there are good theoretical reasons to plead that the decays
$B^0 \to (\rho^0,\omega) + \gamma$ and
$B \to X_d + \gamma$ are well suited to determine the CKM parameters.
We take up these issues in this status report.
%
%

\section{Estimates of $\BBGAMAXS$ in the SM}
\label{sec:effham}
The framework that is used generally to discuss the decays $\BGAMAXS$
is that of an effective theory with
five quarks, obtained by integrating out the
heavier degrees of freedom,
which in the standard model are the top quark and the $W$-boson.
A complete set of dimension-6 operators relevant for the processes
\bsgam  ~and \bsggam ~is contained in the effective Hamiltonian
\begin{equation}
\label{heff}
H_{eff}(b \to s \gamma)
       = - \frac{4 G_{F}}{\sqrt{2}} \, \lambda_{t} \, \sum_{j=1}^{8}
C_{j}(\mu) \, O_j(\mu) \quad ,
\end{equation}
where
$G_F$ is the Fermi constant
coupling constant,
$C_{j}(\mu) $ are the Wilson coefficients evaluated at the scale $\mu$,
and $\lambda_t=V_{tb}V_{ts}^*$ with $V_{ij}$ being the
CKM matrix elements. The overall multiplicative factor $\lambda_t$ follows
from the CKM unitarity and neglecting $\lambda_u$.
The operators $O_j$ read
\bea
\label{operators}
O_1 &=& \left( \bar{c}_{L \b} \g^\m b_{L \a} \right) \,
        \left( \bar{s}_{L \a} \g_\m c_{L \b} \right)\,, \nonumber \\
O_2 &=& \left( \bar{c}_{L \a} \g^\m b_{L \a} \right) \,
        \left( \bar{s}_{L \b} \g_\m c_{L \b} \right) \,,\nonumber \\
O_3 &=& \left( \bar{s}_{L \a} \g^\m b_{L \a} \right) \, \left[
        \left( \bar{u}_{L \b} \g_\m u_{L \b} \right) + ... +
        \left( \bar{b}_{L \b} \g_\m b_{L \b} \right) \right] \,,
        \nonumber \\
O_4 &=& \left( \bar{s}_{L \a} \g^\m b_{L \b} \right) \, \left[
        \left( \bar{u}_{L \b} \g_\m u_{L \a} \right) + ... +
        \left( \bar{b}_{L \b} \g_\m b_{L \a} \right) \right] \,,
        \nonumber \\
O_5 &=& \left( \bar{s}_{L \a} \g^\m b_{L \a} \right) \, \left[
        \left( \bar{u}_{R \b} \g_\m u_{R \b} \right) + ... +
        \left( \bar{b}_{R \b} \g_\m b_{R \b} \right) \right] \,,
        \nonumber \\
O_6 &=& \left( \bar{s}_{L \a} \g^\m b_{L \b} \right) \, \left[
        \left( \bar{u}_{R \b} \g_\m u_{R \a} \right) + ... +
        \left( \bar{b}_{R \b} \g_\m b_{R \a} \right) \right] \,,
        \nonumber \\
O_7 &=& (e/16\p^{2}) \, \bar{s}_{\a} \, \sigma^{\m \n}
      \, (m_{b}(\mu)  R + m_{s}(\mu)  L) \, b_{\a} \ F_{\m \n} \,,
        \nonumber \\
O_8 &=& (g_s/16\p^{2}) \, \bar{s}_{\a} \, \sigma^{\m \n}
      \, (m_{b}(\mu)  R + m_{s}(\mu)  L) \, (\l^A_{\a \b}/2) \,b_{\b}
      \ G^A_{\m \n} \quad ,
        \nonumber \\
\eea
where $e$ and $g_s$ are the electromagnetic and the strong
coupling constants, respectively. In the magnetic moment type
operators $O_7$ and $O_8$, $F_{\m \n}$ and $G^A_{\m \n}$
denote the electromagnetic and the gluonic field strength
tensors, respectively. The subscripts on the quark fields
$L\equiv (1-\g_5)/2$ and $R\equiv (1+\g_5)/2$
denote the left and right-handed projection operators, respectively.
QCD corrections to the decay rate for $b \to s \g$
bring in large logarithms of the form $\a_s^n(m_W) \, \log^m(m_b/M)$,
where $M=m_t$ or $m_W$ and $m \le n$ (with $n=0,1,2,...$).
Using the renormalization group equations the Wilson coefficient
can be calculated at the scale $\mu \approx m_b$ which
is the relevant scale for $B$ decays.
To leading logarithmic precision,
it is sufficient to know the leading order
anomalous dimension matrix and the matching
$C_i(\m=m_W)$ to lowest order (i.e., without
QCD corrections) \cite{InamiLim}.
The $8 \times 8$ anomalous dimension matrix is given in
\cite{Ciuchini}, from where references to earlier calculations can also
be obtained, the
Wilson coefficients
are explicitly listed in
\cite{Buras94} and the
numerical values of these
coefficients being used here can be seen in \cite{ag95}.

\indent
It has become customary to calculate the
branching ratio for the radiative decay $\BGAMAXS$ in terms of the
semileptonic decay branching ratio $\BR (B \to X\ell \nu_\ell)$
\begin{equation}
\label{brdef}
\BR ( B \ra  X_{s} \g) = [\frac{\Gamma(B \ra
 X_{s} + \gamma)}{\Gamma_{sl}}]
\, R(m_b,\mu) \,\BR (B \to X\ell \nu_\ell) \qquad ,
\end{equation}
where, in the approximation of including the leading-order QCD correction,
 $\G_{sl}$ is given by the expression
\begin{equation}
\label{widthsl}
\G_{sl} = \frac{G_F^2 \, |V_{cb}|^2}{192 \pi^3} \, m_b^5 \,
g(m_c/m_b) \, ( 1-2/3 \frac{\alpha_s}{\pi} f(m_c/m_b))\quad .
\end{equation}
The phase space function $g(z)$
and the function $f(z)$ due to one-loop QCD corrections can be seen in
\cite{Alipiet}.
The radiative decay rate $\Gamma(B \to X_s + \gamma)$ are worked out
in \cite{ag1,ag95},
taking into account $O(\alpha_s)$ virtual and bremsstrahlung
corrections. In
calculating the matrix elements in these papers, the on-shell
subtraction prescription for the quark masses has been used.
Due to the explicit
factors of the running quark masses in the operators $O_7$ and
$O_8$, the $m_b^5$-factor contained in the decay rate $\Gamma(B \to X_s +
\gamma)$
is replaced by the following product
\be
m_b^5 \longrightarrow m_b(pole)^3 ~m_b(\mu)^2 \quad ,
\ee
where $m_b(pole)$ and $m_b(\mu)$ denote the pole mass and the $\overline
{\mbox{MS}}$-running
mass of the $b$ quark, respectively. Since, in the leading order
in $\alpha_s$, the semileptonic decay width $\Gamma_{sl}$ depends on the
product $m_b(pole)^5$, the ratio of the two decay widths brings in
the correction factor $R(m_b,\mu)$:
\be
\label{rfactor}
R(m_b,\mu)=[m_b(\mu)/m_b(pole)]^2 \quad ,
\ee
as also remarked in \cite{Buras94}. At the one-loop level, these masses are
related:
\be
\frac{\mb(\mu)}{m_b(pole)}=[\frac{\alpha_s(\mu)}{\alpha_s(\mb)}]^{4/\beta_0}
     [1-\frac{4}{3} \frac{\alpha_s(\mu)}{\pi}],
\label{mbrun}
\ee
where $\beta_0=23/3$.
 The parameters used in estimating
the inclusive rates for $\BBGAMAXS$  are
summarized in table \ref{tabparam}.
\begin{table}
\begin{center}
\begin{tabular}{| c | c | }
\hline
Parameter & Range\\
\hline \hline
$\overline{m_t}$ (GeV) & $170 \pm 11$\\
$\mu$ (GeV) & $5.0^{+5.0}_{-2.5}$ \\
$\Lambda_5$ (GeV) & $0.195 ^{+0.065}_{-0.05}$\\
${\cal B}(B \to X \ell \nu_\ell)$ & $(10.4 \pm 0.4)\%$\\
$m_c/m_b$ & $0.29 \pm 0.02$\\
$m_W$ (GeV) & $80.33$ \\
$\alpha_{\footnotesize{\mbox{QED}}}^{-1}$ & $ 130.0$\\
\hline
\end{tabular}
\end{center}
\caption{Values of the parameters used in estimating the branching
ratio $\BBGAMAXS$ in the standard model.}
\label{tabparam}
\end{table}

 We now discuss
$\BBGAMAXS$ in the standard model and theoretical uncertainties on this
quantity \cite{ag95}.
\begin{itemize}
\item Scale dependence of the Wilson coefficients.
\end{itemize}
The largest theoretical
uncertainty stems from the scale dependence of the Wilson coefficients.
As derived explicitly in \cite{ag95},
the decay rate for $\BGAMAXS$
depends on seven of the eight Wilson coefficients in $H_{eff}(b \to s)$,
once one takes into account the bremsstrahlung corrections and is not
factored in terms of a single (effective) coefficient, namely $C_7^{eff}$,
that one encounters
for the two-body decays $b \to s + \gamma$ \cite{Ciuchini,Buras94}.
Numerical values of the two dominant effective coefficients, $C_7^{eff}$
and $C_8^{eff}$, as one varies $\mu$, the QCD scale  $\Lambda_5$, and
the (running) top quark mass in the $\overline{\mbox{MS}}$-scheme
$\bar{m}_t(m_t)$
in the range given in table 1, are:
\begin{eqnarray}
C_7^{eff} &\equiv & C_7 -\frac{C_5}{3} - C_6 = -0.306 \pm 0.050,
\nonumber\\
C_8^{eff} &\equiv & C_8 + C_5 = -0.146 \pm 0.020 .
\label{c78eff}
\end{eqnarray}
This is the  dominant theoretical error on
$\BBGAMAXS$, contributing about $\pm 35 \%$.
\begin{itemize}
\item
Scale-dependence of $\mb(\mu)$
in the operators $O_7$ and $O_8$.
\end{itemize}
This brings into fore the extra (scale-dependent) multiplicative factor
$R(m_b,\mu)$ for the branching ratio $\BBGAMAXS$, as
discussed above. Intrinsic uncertainties in the concept of the pole mass
due to infrared renormalons suggest that one should express all physical
results in terms of the running masses \cite{renormalons}.
This requires recalculating the
decay rate $\BBGAMAXS$ with the running masses, incorporating resummations
of the kind recently undertaken for the semileptonic $B$ decay rates
\cite{bslresum}. \begin{itemize}
\item Extrinsic errors in $\BBGAMAXS$
\end{itemize}
The next largest
error arises from the parameters which are extrinsic to the decay
$\BGAMAXS$ and have crept in due to
normalizing the
branching ratio $\BBGAMAXS$ in terms of $\BR (B \to X \ell \nu_\ell)$.
The first of  these extrinsic errors is related to the uncertainty
in the ratio  $m_c/m_b$.
Using for the $b$ quark pole mass $m_b(pole)=4.8 \pm 0.15$ GeV
\cite{bqmass} and $m_b-m_c = 3. 40$ GeV \cite{HQET2}, one gets
$m_c/m_b=0.29 \pm 0.02$.
Taking into account the
experimental error of $\pm 4.1 \%$ on $\BR(B \to X \ell \nu_\ell)$
\cite{Gibbons}, one estimates an extrinsic error of
$\pm 12 \%$ on $\BBGAMAXS$.

\indent
Assuming $|V_{ts}|/|V_{cb}=1$ \cite{PDG},
the branching
ratio $\BR(B \to X_s \g)$ calculated
 as a function of the top quark mass is shown in Fig. 1 \cite{ag95}.
\begin{figure}[htb]
\vspace{0.10in}
\centerline{
\epsfysize=3in
\rotate[r]{
\epsffile{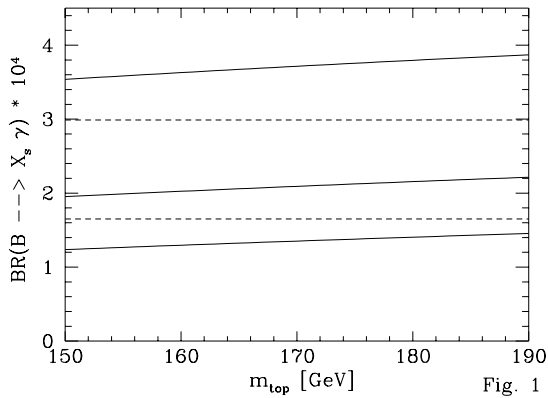}
}
}
\vspace{0.08in}
\caption[]{
$\BR(B \to X_s \g)$ as a function of the $\overline{\mbox{MS}}$ top
quark mass. The
three solid lines correspond to the variation of the parameters
$\mu$ and $\Lambda_5$ as described in the text.
The experimental $(\pm 1\sigma)$-bounds
from CLEO \cite{CLEOrare2} are shown by the dashed lines (from
\protect\cite{ag95}). \label{fig:1}}
\end{figure}
For all three solid curves the quark mass ratio is fixed at
$\mc/\mb=0.29$. The top
solid curve is drawn for $\mu=2.5$ GeV and $\Lambda_5 = 0.260$ GeV.
The bottom solid curve is for $\mu=10$ GeV and $\Lambda_5 = 0.145$
GeV,
and the  middle solid curve corresponds to the central values of
the input parameters in table 1.
Using $\overline{m}_t=(170 \pm 11)$ GeV, and adding the extrinsic error,
one obtains:
\be
\BBGAMAXS=(2.55 \pm 1.28) \times 10^{-4} \,,
\ee
to be compared with the
CLEO measurement $\BBGAMAXS = (2.32 \pm 0.67) \times 10^{-4}$.
The $(\pm 1\sigma)$-upper and -lower bound from the CLEO measurement
are shown in Fig. 1 by dashed lines. The agreement between SM
and experiment is good, given the large uncertainties on both.
In \cite{ag95}, the branching ratio $\BBGAMAXS$ has been calculated as a
function
of the CKM matrix element ratio squared $(|V_{ts}|/|V_{cb}|)^2$, varying
$\bar{m}_t$, $\mu$ and $\Lambda_5$ in the range specified in table 1.
Using the $(\pm 1 \sigma)$-experimental bounds on $\BBGAMAXS$,
 one infers \cite{ag95}:
\be
 |V_{ts}|/|V_{cb}|=1.10 \pm 0.43 \,,
\ee
which is consistent with the indirect constraints from the CKM
unitarity \cite{PDG} yielding $ |V_{ts}|/|V_{cb}| \simeq 1.0$
 but imprecise. Further
improvements require reducing the perturbative scale($\mu$)-dependence of
the decay rate, which in turn implies calculations of the next-to-leading
order terms, and more accurate measurements.
%

\section{Photon Energy Spectrum in $\BGAMAXS$}
The two-body partonic process $b \to s \gamma$
yields a photon energy spectrum $1/\Gamma d \Gamma (b \to s
\gamma) = \delta(1-x)$, where
the scaled photon energy $x$ is defined as
$ E_\g = (m_b^2-m_s^2)/(2 \, m_b )\, x $;
$x$ then varies in the interval $[0,1]$.
Perturbative QCD corrections, such as
$b \to s \gamma + g$, give a characteristic bremsstrahlung spectrum
peaking near the end-points, $E_\gamma \to
E_\gamma ^{max}$ and $E_\gamma \to 0$, arising from the soft-gluon and
soft-photon configurations, respectively. As long as the $s$-quark mass
is non-zero, there is no collinear singularity in the spectrum.
Near the end-points, one
has to improve the spectrum obtained in fixed order perturbation
theory. This is usually done by isolating and exponentiating
the leading behaviour in $\alpha_{em}\alpha_s(\mu)^m
\log^n (1-x)$ and $\alpha_{em}\alpha_s(\mu)^m \log^n x$, with $m\leq n$,
where $\mu$ is a typical momentum in the decay $\BGAMAXS$. The running of
$\alpha_s$ is a non-leading effect, but  as it is characteristic of QCD it
modifies the Sudakov-improved end-point photon energy spectrum \cite{KS94}
compared to its  analogue in QED \cite{Sudakov}.
Away from the end-points, the photon energy
spectrum has to be calculated completely in a given order in $\alpha_s$ in
perturbation theory \cite{ag1,ag95}.

The complete photon energy spectrum in $\BGAMAXS$ is at present
not calculable in QCD from first principles. The situation is very much
analogous to that of other hadronic structure functions.
It has
been observed in a number of papers
\cite{neubertbsg,Bigietal,KS94}, that the x-moments
of the inclusive photon energy spectrum in $\BGAMAXS$ and those of the
lepton energy spectrum in the decay $B \to X_u \ell \nu_\ell$ are related.
Defining the moments as:
\bea
 \M_n(\BGAMAXS) &\equiv & \frac{1}{\Gamma} \int_{0}^{M_B/m_b} dx x^{n-1}
                   \frac{d \Gamma}{dx} \\ \nonumber
\M_n (B \to X_u \ell \nu_\ell) &\equiv&  - \int_{0}^{M_B/m_b} dx x^n
\frac{d}{dx}\big(\frac{1}{\Gamma_\ell}\frac{d \Gamma_\ell}{dx}\big) \\
\nonumber
&=& \frac{n}{\Gamma_\ell} \int_{0}^{M_B/m_b} dx x^{n-1} \frac{d
\Gamma_\ell}{dx}~, \label{moments}
\eea
the ratios of the moments are free of non-perturbative complications.
The moments $\M_n$ have been worked out in
the leading non-trivial order in perturbation theory and
the results can be expressed as:
\be
\M_n \sim 1 + \frac{\alpha_s}{2 \pi} C_F(A\log^2n + B \log n + \mbox{const.})
\ee
where $C_F=4/3$,
the leading coefficient is universal with $A=-1$ \cite{Sudakov}, and the
non-leading coefficients are process dependent;
$B=7/2$ \cite{ag1} and $B
=31/6$ \cite{KJ},
for $\BGAMAXS$ and $B \to X_u \ell \nu_\ell$, respectively. Measurements of
the moments could eventually be used to relate the CKM matrix element
$V_{ts}$ and
$V_{ub}$. That this method will give competitive values for $V_{ub}$,
however, depends on whether or not the coefficient functions
in $\GGAMAXS$ discussed
in the previous section are known to the desired level of
theoretical accuracy.

  We shall leave such theoretically improved
 comparisons for future Rencontres de Moriond and
confine ourselves
to the discussion of the present state-of-the-art
comparison of the measured photon energy spectrum in $\BGAMAXS$
 with the perturbative QCD-improved treatment of the same.
The analysis that we discuss here \cite{ag1,ag95} treats the
 non-perturbative effects
 in terms of a $B$-meson wave function. In this model
\cite{Alipiet}, which admittedly is simplistic but not necessarily wrong,
  the $b$ quark in $B$ hadron is assumed to have a Gaussian
distributed Fermi motion determined by a non-perturbative parameter, $p_F$,
\begin{equation}
\label{lett13}
 \phi(p)= \frac {4}{\sqrt{\pi}{p_F}^3} \exp (\frac {-p^2}{{p_F}^2})
\quad , \quad p = |\vec{p}|
\end{equation}
with the wave function normalization
$ \int_0^\infty \, dp \, p^2 \, \phi(p) = 1.$
The photon energy spectrum from
the decay of the $B$-meson at rest is then given by
\begin{equation}
\label{lett15}
 \frac{d\Gamma}{dE_\gamma}= \int_0^{p_{max}} \, dp \, p^2 \, \phi(p)
  \frac {d\Gamma_b}{dE_\gamma}(W,p,E_\g) \quad ,
\end{equation}
where $p_{max}$ is the maximally allowed value of $p$ and
$ \frac{d\Gamma_b}{dE_\g}$
 is the photon energy spectrum from the decay of the $b$-quark in
flight, having a momentum-dependent mass $W(p)$.

An analysis of the CLEO photon
energy spectrum has been undertaken in \cite{ag95}  to determine
the non-perturbative parameters of this model, namely
$m_b(pole)$ and  $p_F$.
The experimental
errors are still large and the fits result in relatively small $\chi^2$
values; the minimum, $\chi^2_{min}=0.038$, is obtained
for $p_F=450$ MeV and
$m_b(pole)=4.77$ GeV, in good agreement with theoretical estimates of
the same, namely
$m_b(pole)= 4.8\pm 0.15$ GeV \cite{bqmass} and
$p_F^2=\mu_\pi^2/2= 0.25 \pm 0.05$ GeV$^2$
obtained from the QCD sum rules \cite{BB94}.
In Fig. 2 we have plotted the photon energy spectrum normalized
to unit area in the interval between 1.95 GeV and 2.95 GeV for
the parameters which correspond to the minimum $\chi^2$ (solid curve)
and for another set of parameters that lies near the
$\chi^2$-boundary defined by $\chi^2=\chi^2_{min} +1$.
(dashed curve). Data from CLEO \cite{CLEOrare2} are also
shown. Further details of this analysis can be seen in \cite{ag95}.
 \begin{figure}[htb]
\vspace{0.10in}
\centerline{
\epsfysize=3in
\rotate[r]{
\epsffile{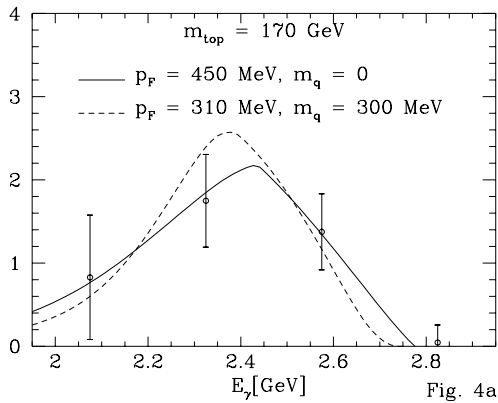}
}
}
\vspace{0.08in}
\caption[]{Comparison of the normalized photon energy distribution
using the
 CLEO data \protect\cite{CLEOrare2} corrected for detector effects and
theoretical
distributions from \protect\cite{ag95} , both  normalized to unit area in
the photon energy interval between 1.95 GeV and 2.95 GeV. The solid
 curve corresponds to the values with the minimum $\chi^2$,
 $(m_q,p_F)$=(0,450 MeV), and the dashed curve to the values
 $(m_q,p_F)$=(300 MeV, 310 MeV).
\label{fig:4}}
\end{figure}

\section{Inclusive radiative decays \bgamaxd }
\indent
The theoretical interest in the standard model in the
(CKM-suppressed) inclusive radiative decays
\bgamaxd\ lies in the first place in the
 possibility of determining the CKM-Wolfenstein
parameters $\rho$ and $\eta$ \cite{Wolfenstein}.
 The relevant region in the
decays $B \to X_d + \gamma$ is the end-point photon energy spectrum,
which has to be measured requiring that
 the hadronic system $X_d$ recoiling against the
photon does not contain strange hadrons to suppress the large-$E_\g$
photons from the decay $\BGAMAXS$. Assuming that this is feasible,
one can determine
 from the ratio of the decay rates
$\BBGAMAXD/\BBGAMAXS$ the CKM-Wolfenstein parameters.
 This measurement was proposed in \cite{ag2}, where
the final-state spectra were also worked out.

\indent
 In close analogy
with the \bgamaxs\ case discussed earlier,
the complete set of dimension-6 operators relevant for
the processes $b \to d \gamma$ and $b \to d \gamma g$
can be written as:
\begin{equation}
\label{heffd}
H_{eff}(b \to d)=
 - \frac{4 G_{F}}{\sqrt{2}} \, \xi_{t} \, \sum_{j=1}^{8}
C_{j}(\mu) \, \hat{O}_{j}(\mu),\quad
\end{equation}
where $\xi_{j} = V_{jb} \, V_{jd}^{*}$ for $j=t,c,u$. The operators
 $\hat{O}_j, ~j=1,2$, have implicit in them CKM factors. In the
Wolfenstein parametrization \cite{Wolfenstein}, one can express these
factors as :
$\xi_u = A \, \lambda^3 \, (\rho - i \eta),
{}~\xi_c = - A \, \lambda^3 ,
{}~\xi_t=-\xi_u - \xi_c$.
We note that all three CKM-angle-dependent quantities
$\xi_j$ are of the
same order of magnitude, $O(\lambda^3)$, where $\lambda =\sin \theta_C
\simeq 0.22$.
This is an important difference as compared to the effective
Hamiltonian ${\cal H}_{eff}(b \to s)$ written earlier,
 in which case the
 effective Hamiltonian
 factorizes into an overall CKM factor $\lambda_t$.
For calculational ease,  this difference
can be implemented by defining the operators $\hat{O}_1$ and $\hat{O}_2$
entering in $H_{eff}(b \to d)$ as follows \cite{ag2}:
\begin{eqnarray}
\label{basis}
&&\hat{O}_{1} =
 -\frac{\xi_c}{\xi_t}(\bar{c}_{L \beta} \go{\mu} b_{L \alpha})
(\bar{d}_{L \alpha} \gu{\mu} c_{L \beta})
 -\frac{\xi_u}{\xi_t}(\bar{u}_{L \beta} \go{\mu} b_{L \alpha})
(\bar{d}_{L \alpha} \gu{\mu} u_{L \beta}) ,\nonumber \\
&& \hat{O}_{2} =
-\frac{\xi_c}{\xi_t}(\bar{c}_{L \alpha} \go{\mu} b_{L \alpha})
(\bar{d}_{L \beta} \gu{\mu} c_{L \beta})
 -\frac{\xi_u}{\xi_t}(\bar{u}_{L \alpha} \go{\mu}
b_{L \alpha}) (\bar{d}_{L \beta} \gu{\mu} u_{L \beta}) ,
\end{eqnarray}
and the rest of the operators $(\hat{O}_j;~j=3...8)$ are
defined like their
counterparts ${O}_j$ in $H_{eff}(b \to s)$, with the obvious replacement
$s \to d$. With this definition, the matching conditions $C_j(m_W)$
 and the solutions
of the RG equations yielding $C_j(\mu)$ become
identical for the two operator basis $O_j$ and $\hat{O}_j$.
 It has been explicitly checked in the $O(\alpha_s)$ calculations
of the decay rate and photon energy spectrum  involving $b \to d g$ and
$b \to d g \gamma$ transitions that the limit $m_u \to 0$
for the decay rate $\GGAMAXD$
 exists \cite{ag2}.  From this it
follows that, in the leading order QCD corrections, there are no
logarithms of the type $\alpha_s \log (m_u^2/m_c^2)$ \cite{Ricciardi}.
Some papers, estimating LD-contributions in radiative $B$ decays, seem
to contradict this by assuming light-quark contributions which have such
spurious log-dependence. There is no calculational basis for this
assumption.
 In higher orders, such terms must be absorbed in the
non-perturbative functions.
 On the other hand, as far as the dependence of the decay rate and spectra
on the external light quark masses is concerned,
one encounters logarithms of the
type $\alpha_{em}\alpha_s (1+(1-x)^2)/x \log (m_b^2/m_s^2)$
 (for $b \to s g \gamma)$ and
$\alpha_{em}\alpha_s (1+(1-x)^2)/x \log (m_b^2/m_d^2)$ (for $b \to d g
\gamma$)
 near the soft-photon
($x \to 0)$ region \cite{ag95}, which can, however, be exponentiated
\cite{klp95}.
The essential difference between  $\GGAMAXS$ and $\GGAMAXD$
lies in the matrix elements of the first two operators $O_1$ and $O_2$
(in $H_{eff}(b \to s)$) and $\hat{O}_1$ and $\hat{O}_2$ (in $H_{eff}(b
\to d)$).
The derivation of the inclusive decay rate and the final-state distributions
in \bgamaxd\ otherwise
goes along very similar lines as for the decays \bgamaxs\ .
The branching ratio  $\BBGAMAXD$
 in the SM  can be written as:
\begin{eqnarray}
\label{branstruc}
&& \BBGAMAXD = D_1 |\xi_t|^2 \nonumber \\
&&                     \{
1 - \frac{1-\rho}{(1-\rho)^2 + \eta^2} \, D_2
  - \frac{\eta}{(1-\rho)^2 + \eta^2} \, D_3
 + \frac{D_4}{(1-\rho)^2 + \eta^2} \} , \quad
\end{eqnarray}
where the functions $D_i$ depend on the parameters listed in table 1. The
uncertainty on this branching ratio from the parametric dependence is very
similar to the one worked out for $\BBGAMAXS$.
 For the central values of the parameters in table 1, one gets	:
$D_1=0.21, ~D_2=0.17, ~D_3=0.03, ~D_4=0.10$.
To get the inclusive branching ratio
the CKM parameters $\rho$ and $\eta$ have to be constrained from the
unitarity fits. Taking the parameters from a recent fit, one gets
$ 5.0 \times 10^{-3} \leq \vert \xi_t  \vert \leq 1.4 \times
10^{-2}$
(at 95\% C.L.) \cite{al95},
yielding an order of magnitude uncertainty in $\BBGAMAXD$ -
hence the interest in measuring it. Taking the central values of the
fit parameters
$A=0.8, \lambda=0.2205, \eta =0.34$ and $\rho=-0.07$ \cite{al95},
 one gets $\BBGAMAXD = (1.7\pm 0.85) \times 10^{-5}$, which is
approximately a factor 10 -20
smaller than the CKM-allowed branching ratio $\BBGAMAXS$, measured by CLEO
\cite{CLEOrare1}.
%

\vspace*{3.0ex}
\section{ Estimates of ${\cal B}(B \to V + \gamma )$
 and Constraints on the CKM Parameters $\rho$ and $\eta$}
\indent
Exclusive radiative
 $B$ decays $B \to V + \gamma$, with $V=K^*,\rho,\omega$, are also
potentially
very interesting from the point of view of determining the CKM parameters
\cite{abs93}. The extraction of these parameters would, however,  involve a
trustworthy
estimate of the SD- and LD-contributions in the decay amplitudes.
\par
  The SD-contribution in the
 exclusive decays $(B_u, B_d) \to (K^*,\rho) + \gamma$,
$B_d \to \omega + \gamma$  and the
corresponding $B_s$ decays, $B_s \to (\phi,K^*) + \gamma $,
involve the magnetic moment operator $O_7$ and the related one obtained
by the obvious change $s \to d$, $\hat{O}_7$.
The transition form factors governing the radiative $B$ decays
 $B \to V + \gamma$ can be generically  defined as:
\be
 \langle V,\lambda |\frac{1}{2} \bar \psi \sigma_{\mu\nu} q^\nu b
 |B\rangle  =
     i \epsilon_{\mu\nu\rho\sigma} e^{(\lambda)}_\nu p^\rho_B p^\sigma_V
F_S^{B\rightarrow V}(0).
\label{defF}
\ee
Here $V$ is a vector meson
with the polarization vector $e^{(\lambda)}$,
$V=\rho, \omega, K^*$ or $\phi$;
$B$ is a generic
$B$-meson $B_u, B_d$ or $B_s$, and $\psi$ stands for the
field of a light $u,d$ or $s$ quark. The vectors $p_B$, $p_V$ and
$q=p_B-p_V$
correspond to the 4-momenta of the initial $B$-meson and the
outgoing vector
meson and photon, respectively. In (\ref{defF}) the QCD
renormalization of the $\bar \psi \sigma_{\mu\nu} q^\nu b$ operator
is implied.
 Keeping only the SD-contribution
 leads to obvious relations among the exclusive
decay rates, exemplified here by the decay
rates for $(B_u,B_d) \to \rho + \gamma$ and $(B_u,B_d) \to K^* + \gamma$:
\be
\frac{\Gamma ((B_u,B_d) \to \rho + \gamma)}
     {\Gamma ((B_u,B_d) \to K^* + \gamma)}
  = \frac{\vert \xi_t \vert^2}{\vert\lambda_t \vert ^2}
      \frac{\vert F_S^{B \to \rho }(0)\vert^2}
          {\vert F_S^{B \to K^* }(0)\vert^2} \Phi_{u,d}
  =\kappa_{u,d}[\frac{\absvtd}{\absvts}]^2 \,,
\label{SMKR}
\ee
where $\Phi_{u,d}$ is a phase-space factor which in all cases is close to 1
and $\kappa_{i} \equiv [F_S^{B_i \to \rho \gamma}/F_S^{B_i \to K^*
\gamma}]^2$ is the ratio of the (SD) form factors squared.
The transition form factors $F_s$
 are model dependent. However, their ratios, i.e. $\kappa_i$,
 should be more reliably calculable as they depend
essentially only on the SU(3)-breaking effects.
 If the SD-amplitudes were the only contributions, the measurements of the
 CKM-suppressed radiative decays $(B_u,B_d) \to \rho + \gamma ,
{}~B_d \to \omega + \gamma$ and $B_s \to K^* + \gamma$ could be
used in conjunction with the decays $(B_u,B_d) \to K^* + \gamma$ to
determine
the CKM parameters. The present experimental upper limits on the CKM ratio
$\absvtd/\absvts$ from radiative $B$ decays
are indeed based on this assumption, yielding \cite{cleotdul}:
\begin{equation}
\frac{\vert V_{td} \vert }{\vert V_{ts}} \vert \leq 0.75~,
\end{equation}
with a theoretical dispersion estimated in the range $0.64$ - $0.75$,
depending on the models used for the $SU(3)$ breaking effects
in the form factors \cite{abs93,SU3f2}.

  The possibility of significant
LD- contributions in
radiative $B$ decays from the light quark intermediate states
has been raised in a number of papers
\cite{ldall}.
Their amplitudes necessarily involve other CKM matrix elements and hence the
simple factorization of the decay rates in terms of the CKM factors
involving $\absvtd$ and $\absvts$ no longer holds thereby
 invalidating the relationships
given above. In what follows, we argue that the CKM-analysis of
 charged $B$-decays,
$B^\pm \to \rho^\pm \gamma$, would require modifications due to the
LD-contributions but the corresponding analysis of the
neutral $B$-decays $B \to (\rho^0,\omega) \gamma$ remains essentially
unchanged.

The LD-contributions in $B \to V + \gamma$  are
induced by the matrix elements of the
four-Fermion operators $\hat{O}_1$ and $\hat{O}_2$ (likewise $O_1$ and
$O_2$). Estimates of these contributions
 require non-perturbative methods.
 This problem has been investigated recently
in \cite{ab95,wyler95} using
 a technique
which treats the photon emission from the light quarks in a theoretically
consistent and model-independent way. This has been combined
with the light-cone QCD sum rule approach to calculate both the SD and LD
--- parity conserving and parity violating --- amplitudes
in the decays $B_{u,d} \to \rho(\omega) + \gamma$.
To illustrate this, we concentrate on the $B_u^\pm$ decays,
$B_u^\pm \to \rho^\pm + \gamma$ and take up the neutral $B$ decays
$B_d \to \rho (\omega) + \gamma$ at the end.
The LD-amplitude of the four-Fermion operators $\hat{O}_1$, $\hat{O}_2$
is dominated by  the
 contribution of the weak annihilation
of valence quarks in the $B$ meson. It is color-allowed for the
decays of charged $B^\pm$ mesons, as shown in fig. 3, where also the
tadpole diagram is shown, which, however, contributes only in the
presence of gluonic corrections, and hence neglected.
In the factorization approximation, one may write the dominant contribution
in the operator $\hat{O_2}$ (here $O^\prime_2$ is the part of $\hat{O_2}$
with the CKM factor $\xi_u/\xi_t$)
\begin{equation}\label{factor}
\langle \rho\gamma | O^\prime_2|B\rangle =
\langle \rho | \bar d \Gamma_\mu u|0\rangle
\langle \gamma | \bar u \Gamma^\mu b|B\rangle
+
\langle \rho\gamma | \bar d \Gamma_\mu u|0\rangle
\langle 0 | \bar u \Gamma^\mu b|B\rangle ~,
\end{equation}
and make use of the definitions of the decay constants
\begin{eqnarray}
\langle 0 | \bar u \Gamma_\mu b|B\rangle  & =& i p_\mu f_B,
\nonumber\\
\langle \rho | \bar d \Gamma_\mu u|0\rangle &=&
\varepsilon^{(\rho)}_\mu m_\rho f_\rho,
\end{eqnarray}
to reduce the problem at hand to the calculation of simpler form factors
induced by vector and axial-vector currents.
\begin{figure}[htb]
\vspace{0.10in}
\vspace*{-3cm}
\centerline{
\epsfysize=6in
{
\epsffile{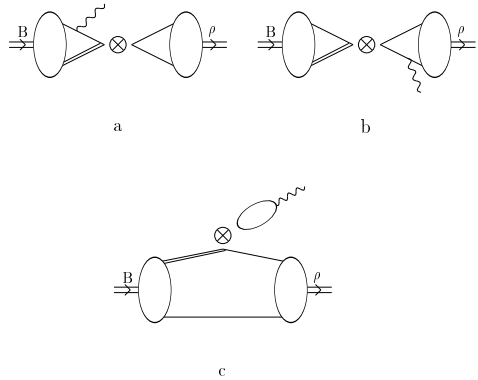}
}
}
\vspace{0.08in}
\vspace*{-8cm}
\caption[]{
 Weak annihilation contributions in $B_u \to \rho  \gamma$
involving the operators $O^\prime_1$ and $O^\prime_2$ denoted by
$\bigotimes$ with the
photon emission from a) the loop containing the $b$ quark, b) the loop
containing the light quark, and c) the tadpole which contributes only
with additional gluonic corrections.
\label{fig:3}}
\end{figure}

The factorization approximation assumed in \cite{ab95,wyler95} has been
tested
(to some extent) in two-body and quasi-two body non-leptonic $B$ decays
involving the transitions $b \to c \bar{c} s$ and $b \to c \bar{u} d$.
It has not been tested experimentally in
radiative $B$ decays. From a theoretical point of view,
 non-factorizable
contributions belong to either the $O(\alpha_s)$ (and higher order)
radiative corrections or to contributions of higher-twist operators to
the sum rules. Their inclusion should not change the conclusions
substantially.

The LD-amplitude in the decay $B_u \to \rho^\pm + \gamma$ can be written
in terms of the form factors $F_1^L$ and $F_2^L$,
\begin{eqnarray}\label{Along}
{\cal A}_{long} &=&
-\frac{e\,G_F}{\sqrt{2}} V_{ub}V_{ud}^\ast
\left( C_2+\frac{1}{N_c}C_1\right) m_\rho
\varepsilon^{(\gamma)}_\mu \varepsilon^{(\rho)}_\nu
\nonumber\\&&{}\times
 \Big\{-i\Big[g_{\mu\nu}(q\cdot p)- p_\mu q_\nu\Big] \cdot 2 F_1^{L}(q^2)
  +\epsilon_{\mu\nu\alpha\beta} p^\alpha q^\beta
 \cdot 2 F_2^{L}(q^2)\Big\}\,.
\label{ratio2}
\end{eqnarray}
Again, one has to invoke a model to calculate the form factors. Estimates
 from the light-cone QCD sum rules give
\cite{ab95}:
\begin{equation}\label{result}
 F^L_1/F_S = 0.0125\pm 0.0010\,,\quad F^L_2/F_S = 0.0155\pm 0.0010 ~,
\end{equation}
where the errors correspond to the variation of the
Borel parameter in the QCD sum rules. Including other possible
uncertainties,
 one expects an accuracy of the ratios in (\ref{result}) of order 20\%.
Since the parity-conserving and parity-violating amplitudes turn out
to be close to each other, $F_1^L\simeq F^L_2 \equiv F_L$,
the ratio of the LD- and the SD- contributions reduces to a number
\begin{equation}\label{ratio2p}
{\cal A}_{long}/{\cal A}_{short}=
R_{L/S}^{B_u\to\rho\gamma}
\cdot\frac{V_{ub}V_{ud}^\ast}{V_{tb}V_{td}^\ast} ~.
\end{equation}
Using $C_2=1.10$, $C_1=-0.235$, $C_7^{eff}=-0.306$ (at the scale $\mu=5$ GeV)
 \cite{ag95} gives:
\begin{equation}\label{result2}
R_{L/S}^{B_u\to\rho\gamma} \equiv
 \frac{4 \pi^2 m_\rho(C_2+C_1/N_c)}{m_b C_7^{eff}}
\cdot\frac{F_L^{B_u \to \rho \gamma}}{F_S^{B_u \to \rho \gamma}}=-0.30\pm
0.07 ~. \end{equation}
 To get a ball-park estimate of the ratio
${\cal A}_{long}/{\cal A}_{short}$, we take the central values of
the CKM matrix elements,
 $V_{ud}=0.9744\pm 0.0010$ \cite{PDG},
$|V_{td}|=(1.0\pm 0.2)\times 10^{-2}$,
$|V_{cb}|=0.039\pm 0.004$ and $|V_{ub}/V_{cb}|=0.08\pm 0.02$ \cite{al95},
 yielding,
\begin{equation}
|{\cal A}_{long}/{\cal A}_{short}|^{B_u\to\rho\gamma}
= |R_{L/S}^{B_u\to\rho\gamma}|
\frac{|V_{ub}V_{ud}|}{|V_{td}V_{bt}|} \simeq 10\% ~.
\end{equation}
The analogous LD-contributions to the neutral $B$ decays
$B_d\to\rho\gamma $ and $B_d\to\omega\gamma $ are
expected to be much smaller, a point
that has also been noted in the context of the VMD and quark model
based estimates \cite{ldall}. In the present approach,
 the corresponding form factors for the decays
$B_d \to \rho^0(\omega)  \gamma$ are obtained from
the ones for the decay $B_u\to\rho^\pm \gamma$ discussed above by the
replacement of the light quark charges
 $e_u\to e_d$, which gives the factor $-1/2$; in addition,
and more importantly, the
LD-contribution to the neutral $B$ decays
is colour-suppressed, which reflects itself
through the replacement of the factor
$a_1\equiv C_2+C_1/N_c$ in (\ref{ratio2}) by
$a_2\equiv C_1+C_2/N_c$. This yields for the ratio
\begin{equation}
\frac{R_{L/S}^{B_d\to\rho\gamma}}{R_{L/S}^{B_u\to\rho\gamma}}=
\frac{e_d a_2}{e_u a_1} \simeq -0.13 \pm 0.05 ,
\end{equation}
where the numbers are based on using
$a_2/a_1 = 0.27 \pm 0.10$ \cite{BHP93}. This would then yield at most
$R_{L/S}^{B_d\to\rho\gamma} \simeq R_{L/S}^{B_d\to\omega\gamma}=0.05$,
which in turn gives
 ${\cal A}_{long}^{B_d\to\rho\gamma}/{\cal A}_{short}^{B_d\to\rho\gamma}
\leq 0.02$. Even if this underestimates the LD-contribution by a factor
2, due to the approximations made in \cite{ab95,wyler95},
it is quite safe to neglect
the LD-contribution in the neutral $B$-meson radiative decays.

The ratio of the CKM-suppressed and CKM-allowed
 decay rates  for charged $B$ mesons
 gets modified due to the LD contributions. Following \cite{GP95},
we ignore the LD-contributions in $\Gamma(B \to K^*\gamma)$. The ratio of
the decay rates in question can therefore be written as:
\begin{eqnarray}\label{ratio3}
\lefteqn{\frac{\Gamma(B_u\to \rho\gamma)}{\Gamma(B_u\to K^*\gamma)}
= \kappa_u \lambda^2[(1-\rho)^2+\eta^2]
}
\nonumber\\&&{}
\times\Bigg\{
1+2\cdot R_{L/S} V_{ud}\frac{\rho(1-\rho)-\eta^2}{(1-\rho)^2+\eta^2}
+(R_{L/S})^2 V_{ud}^2\frac{\rho^2+\eta^2}{(1-\rho)^2+\eta^2}\Bigg\}\,,
\end{eqnarray}
 Using the central value from the estimates of the ratio of the
  form factors squared
$\kappa_u=0.59 \pm 0.08$
  \cite{abs93}, and the presently allowed range of the Wolfenstein parameters
 $\rho$ and
 $\eta$,
 it is shown in \cite{ab95} that
the effect of the LD-contributions is modest but not negligible, introducing
an uncertainty
comparable to the $\sim 15\%$ uncertainty in the overall normalization
due to the $SU(3)$-breaking effects in the quantity $\kappa_u$.

\indent
Neutral $B$-meson radiative decays are less-prone to the LD-effects,
 as argued above, and hence one expects that to a good approximation
the ratio of the decay rates for neutral $B$ meson obtained in the
approximation of SD-dominance remains valid \cite{abs93}:
\begin{equation}
\frac{\Gamma(B_d\to \rho\gamma,\omega\gamma)}{\Gamma(B\to K^*\gamma)}
 = \kappa_d\lambda^2 [(1-\rho)^2+\eta^2]~.
\end{equation}
 Here $\kappa_d$ represents the
 SU(3)-breaking effects in the
form factor ratio squared.
 It is a realistic hope that this relation is
theoretically (almost) on the same footing in the standard model
as the one for the ratio of the $B^0$-$\overline{B^0}$ mixing-induced
mass differences, which satisfies the relation \cite{al95}:
\be
\frac{\delms}{\delmd} = \kappa_{sd}
\left\vert \frac{V_{ts}}{V_{td}} \right\vert^2
= \kappa_{sd}\frac{1}{\lambda^2 [(1-\rho)^2+\eta^2]} ~.
\label{xratio}
\ee
The
 hadronic
uncertainty in this ratio
 is in the SU(3)-breaking factor $\kappa_{sd}\equiv (f_{B_s}^2 \hat{B}_{B_s}/
f_{B_d}^2 \hat{B}_{B_d})$, which involves the pseudoscalar coupling
constants and the so-called bag constants. This quantity is
 estimated as $\kappa_{sd}=1.35 \pm 0.25$ in the QCD sum
rules and lattice QCD approaches. (For details and references, see
\cite{al95}).
The present upper limit for the mass-difference ratio $\delms/\delmd > 12.3
$ at 95 \% C.L. from the ALEPH data \cite{ALEPHxs} provides better
constraint on the CKM parameters, yielding $\absvtd/\absvts < 0.35$
 than the corresponding constraints from the
rare radiative decays $B \to (\rho,\omega) + \gamma$, which give an upper
limit of 0.75 for the same CKM-ratio. We expect experimental sensitivity to
increase in both measurements,
reaching the level predicted for this ratio
in the standard model, $\absvtd/\absvts =0.24 \pm 0.05$ \cite{al95},
in the next several years in the ongoing experiments at CLEO, LEP and
Tevatron, and the forthcoming ones at the $B$ factories and HERA-B.

\indent

 Finally, combining the estimates for the LD- and SD-form factors in
\cite{ab95} and
\cite{abs93}, respectively, and restricting the Wolfenstein
parameters in the range $-0.4 \leq \rho \leq 0.4$ and $ 0.2 \leq \eta
\leq 0.4$, as suggested by the CKM-fits \cite{al95}, we give the
following ranges for the absolute branching ratios:
\begin{eqnarray}\label{ratio4}
{\cal B}(B_u\to \rho\gamma)
&=& (1.9 \pm 1.6) \times 10^{-6} ~,
\nonumber\\
{\cal B}(B_d\to \rho\gamma) &\simeq& {\cal B}(B_d \to \omega \gamma)
= (0.85 \pm 0.65) \times 10^{-6} ~,
\end{eqnarray}
where we have used the experimental value for the branching ratio
${\cal B} (B \to K^* + \gamma) =(4.5 \pm 1.5 \pm 0.9) \times 10^{-5}$
\cite{CLEOrare1},
adding the errors in quadrature. The large error reflects the poor
knowledge of the CKM matrix elements and hence experimental determination
of these branching ratios will put rather stringent constraints on the

Summarizing the effect of the LD-contributions in
radiative $B$ decays, we note that they are
dominantly given by the annihilation diagrams. QCD sum-rule-based
estimates are very encouraging in that they lead to the conclusion
that such contributions are modest in exclusive radiative $B$ decays,
in particular in the neutral $B$-decays $B^0 \to (\rho^0,\omega) + \gamma$.
This estimate should be checked in other theoretically sound frameworks.
Of course, forthcoming data on specific $B$-meson decays will be able to
check this directly.
Presently available data suggest that
 the contribution  of annihilation diagrams in $B$ decays is not significant,
 as seen through the near equality of the lifetimes
for the $B^\pm, ~B_d^0$ and $B_s^0$ mesons and the near equality of
the observed $B^\pm$ and $B^0$ radiative decay rates.
 We have argued that this is very probably
also the case for the CKM-suppressed radiative decays, with $B^\pm \to
\rho^\pm \gamma$ modified by $O(20)\%$ from its SD-rate.

\noindent
{\bf Acknowledgements}:
 I would like to thank Vladimir Braun, Christoph Greub, David London,
and  Hubert Simma
 for numerous helpful
discussions and valuable input. Informative discussions with Arkady
Vainshtein are also gratefully acknowledged.

\vfill
\end{document}